\documentclass[twoside,twocolumn,9pt]{article}
\usepackage{extsizes}
\usepackage[super,sort&compress,comma]{natbib} 
\usepackage[version=3]{mhchem}
\usepackage[left=1.5cm, right=1.5cm, top=1.785cm, bottom=2.0cm]{geometry}
\usepackage{balance}
\usepackage{widetext}
\usepackage{times,mathptmx}
\usepackage{sectsty}
\usepackage{graphicx} 
\usepackage{lastpage}
\usepackage[format=plain,justification=raggedright,singlelinecheck=false,font={stretch=1.125,small,sf},labelfont=bf,labelsep=space]{caption}
\usepackage{float}
\usepackage{fancyhdr}
\usepackage{fnpos}
\usepackage[english]{babel}
\usepackage{array}
\usepackage{droidsans}
\usepackage{charter}
\usepackage[T1]{fontenc}
\usepackage[usenames,dvipsnames]{xcolor}
\usepackage{setspace}
\usepackage[compact]{titlesec}
\usepackage{bm}
\usepackage{amssymb}
\usepackage{notes2bib}

\usepackage{epstopdf}

\definecolor{cream}{RGB}{222,217,201}

\begin{document}

\pagestyle{fancy}
\thispagestyle{plain}
\fancypagestyle{plain}{

}

\makeFNbottom
\makeatletter
\renewcommand\LARGE{\@setfontsize\LARGE{15pt}{17}}
\renewcommand\Large{\@setfontsize\Large{12pt}{14}}
\renewcommand\large{\@setfontsize\large{10pt}{12}}
\renewcommand\footnotesize{\@setfontsize\footnotesize{7pt}{10}}
\makeatother

\renewcommand{\thefootnote}{\fnsymbol{footnote}}
\renewcommand\footnoterule{\vspace*{1pt}%
\color{cream}\hrule width 3.5in height 0.4pt \color{black}\vspace*{5pt}} 
\setcounter{secnumdepth}{5}

\makeatletter 
\renewcommand\@biblabel[1]{#1}            
\renewcommand\@makefntext[1]%
{\noindent\makebox[0pt][r]{\@thefnmark\,}#1}
\makeatother 
\renewcommand{\figurename}{\small{Fig.}~}
\sectionfont{\sffamily\Large}
\subsectionfont{\normalsize}
\subsubsectionfont{\bf}
\setstretch{1.125} 
\setlength{\skip\footins}{0.8cm}
\setlength{\footnotesep}{0.25cm}
\setlength{\jot}{10pt}
\titlespacing*{\section}{0pt}{4pt}{4pt}
\titlespacing*{\subsection}{0pt}{15pt}{1pt}

\fancyfoot{}
\fancyfoot[LO,RE]{\vspace{-7.1pt}}
\fancyfoot[CO]{\vspace{-7.1pt}\hspace{13.2cm}}
\fancyfoot[CE]{\vspace{-7.2pt}\hspace{-14.2cm}}
\fancyfoot[RO]{\footnotesize{\sffamily{1--\pageref{LastPage} ~\textbar  \hspace{2pt}\thepage}}}
\fancyfoot[LE]{\footnotesize{\sffamily{\thepage~\textbar~1--\pageref{LastPage}}}}
\fancyhead{}
\renewcommand{\headrulewidth}{0pt} 
\renewcommand{\footrulewidth}{0pt}
\setlength{\arrayrulewidth}{1pt}
\setlength{\columnsep}{6.5mm}
\setlength\bibsep{1pt}

\makeatletter 
\newlength{\figrulesep} 
\setlength{\figrulesep}{0.5\textfloatsep} 

\newcommand{\topfigrule}{\vspace*{-1pt}%
\noindent{\color{cream}\rule[-\figrulesep]{\columnwidth}{1.5pt}} }

\newcommand{\botfigrule}{\vspace*{-2pt}%
\noindent{\color{cream}\rule[\figrulesep]{\columnwidth}{1.5pt}} }

\newcommand{\dblfigrule}{\vspace*{-1pt}%
\noindent{\color{cream}\rule[-\figrulesep]{\textwidth}{1.5pt}} }

\makeatother


\twocolumn[
\begin{@twocolumnfalse}
\sffamily

\begin{centering}
\vspace{1.3cm}
\LARGE{\textbf{How roughness affects the depletion mechanism}} \\
\vspace{0.4cm}
\large{Pietro Anzini and Alberto Parola$^{\dag}$} \\
\end{centering}
\vspace{0.7cm}
\hspace{0.1\linewidth}\begin{minipage}{0.8\textwidth}
\normalsize{We develop a simple model, in the spirit of the Asakura-Oosawa
theory, able to describe the effects of surface roughness on the depletion potential as a function of a small set of parameters. 
The resulting explicit expressions are easily computed, without free parameters, for a wide range of physically interesting conditions.
Comparison with the recent numerical simulations [M. Kamp et al., \textit{Langmuir}, 2016, \textbf{32}, 1233] shows an encouraging
agreement and allows to predict the onset of colloidal aggregation in dilute suspensions of rough particles.
Furthermore, the model proves to be suitable to investigate the role of the geometry of the roughness.
}
\vspace{1.3cm}
\end{minipage}
\end{@twocolumnfalse} \vspace{0.6cm}
]


\renewcommand*\rmdefault{bch}\normalfont\upshape
\rmfamily
\section*{}
\vspace{-1cm}


\footnotetext{\textit{Dipartimento di Scienza e Alta Tecnologia, Universit\`a degli Studi dell'Insubria, Via Valleggio 11, 22100 Como, Italy}}


\footnotetext{$^{\dag}$~E-mail: alberto.parola@uninsubria.it}


\section{Introduction}
The study of solvent-mediated interactions in colloids dates back to the seminal work by Asakura and Oosawa~\cite{asa_osava_54,asa_osava_58} 
(AO) and the subsequent independent analysis by Vrij~\cite{vrij}: 
two large colloidal particles suspended in a dilute polymer solution suffer an effective attractive interaction 
arising from the depletion of solutes between them. The range of the interaction equals 
the depletant diameter, whereas its strength increases with polymer density.
In their work, Asakura and Oosawa considered a system where the depletant is an ideal gas, 
the colloidal particles are hard spheres and the mutual interaction between colloids and depletant is of the excluded volume type. 
This particular choice of the inter-particle interactions gives rise to a purely entropic potential, 
proportional to the temperature $T$, the polymer fugacity $z_p$ and the overlap volume $V^{ov}$ between the colloidal depletion layers, 
i.e. the spherical shells surrounding the colloidal particles from which  
the depletant particles are excluded. In formulae
\begin{equation}
\beta v_{\mathrm{AO}}(r) =
\begin{cases}
+\infty & \text{if the two particles overlap} \\
-z_p V^{ov}(r) & \text{if the two depletion layers overlap} \\
0 & \text{elsewhere}
\end{cases}
\label{eq:ao_generale}
\end{equation}
where $\beta=1/k_\mathrm{B}T$. 
Even if the original analysis focused the attention on two colloidal spherical particles~\cite{asa_osava_58}, 
it has been shown that the AO result is correct for an arbitrary number of colloidal particles when the ratio $q=\sigma/2R$
between the depletant and colloidal diameter is less than $0.155$~\cite{binary_mapping}. In other words, it is possible to map exactly a binary 
AO-mixture into an effective one component system interacting through the AO pair potential.
Extensive experimental and numerical investigations of the 
effective potential between hard spheres mediated by non-adsorbing polymer coils identified the AO model as the
epitome of the depletion mechanism~\cite{ao_exp_laser,ao_exp_afm}.

A peculiar prediction of the AO model is that the strength of attraction increases with the
size of the colloidal particles, at fixed solute volume fraction, implying that macroscopic
objects should feel extremely large attractive forces at short distances. 
This paradoxical circumstance becomes evident considering the contact value of the AO potential, which 
diverges as the size ratio $q$ tends to zero: 
\begin{equation}
\beta v_{\mathrm{AO}}^{(c)}=-\eta \left(1+\frac{3}{2q}\right).
\label{eq:contAO}
\end{equation}
where $\eta$ is the polymer reservoir packing fraction. 
Such a strong, short ranged, divergence implies that smooth colloidal particles immersed in a 
molecular solvent would stick together due to depletion interactions. This unphysical behaviour, which
contrasts with our daily experience, originates from neglecting the irregularity of any particle surface on
molecular scales and calls for a generalisation of the AO approach to include the effects of surface roughness.

Indeed, it is well known, starting from the studies by Pine and coworkers~\cite{pine_1996,dinismore_1999}, that surface geometry
strongly affects the overlap volumes, leading to the suppression or the enhancement of 
the depletion interaction\cite{dietrich_curved,dietrich_patterned}.
Increasing the overlap volumes reflects in a stronger depletion potential to the point of 
allowing the realisation of site specific interactions between colloidal particles, as theoretically 
predicted and experimentally shown in 
recent works regarding lock and key colloids~\cite{lockandkey_dietrich,lockandkey}.
It has been experimentally shown~\cite{strook_generouh,mason_exp} that depletion attraction 
between rough colloids can be suppressed when the height of the asperities becomes larger 
than the depletant because the overlap volume is significantly reduced by a small amount of surface corrugation. 
This tunable behaviour of the depletion interaction through surface roughness seems to be promising in order to 
control particle aggregation, explore different phases and design novel materials
~\cite{BarryZvonomir_2010,Kraft_2012,irvine_2015,dimpled_crumpled_2016,mickey_mouse_2016,wolters_2017}.
 
Notwithstanding the widespread awareness of the effects of roughness on the depletion mechanism, 
only a limited number of studies dealing with solvent mediated interaction between rough objects can be found in the literature. 
Zhao and Mason~\cite{mason_theo} investigated the problem by computing the minimum of the 
depletion potential between platelets decorated by hemispherical asperities with different height, radius 
and configurations, and their results corroborate previous experimental findings~\cite{mason_exp}.
More recently, Schweizer and collaborators tackled the question of the role of surface topography 
on the depletion interaction through a hybrid Monte Carlo plus integral equation theory approach~\cite{Schweizer}.
They consider the interaction between corrugated ``raspberry'' particles immersed in a hard sphere polymer fluid
for different size ratios and packing fractions, finding that the resulting effective interaction is affected by 
the competition among the standard depletion and the excess entropic contributions arising from the fluid 
present within the surface interstices. The analysis shows that surface corrugation suppresses
the depletion induced aggregation for values of the depletant diameter close to the height of the roughness. 
A recent work \cite{mariolina_2016} evaluates, by means of Monte Carlo simulations, the effective potential 
between two spherical hard colloids, whose surface is decorated with smaller spherical particles,
immersed in an ideal depletant, comparing the results with experiments in a colloidal suspension of silica particles.
An interesting outcome of this work is that the best reduction of the depletion
potential is obtained for incomplete surface coverings and for a depletant with approximately the 
same size as the particles attached to the colloidal surface.

Despite the theoretical efforts devoted to this problem, a simple analytical or semi-analytical approach 
able to capture the physics of the problem is still missing: In particular it would be interesting 
to determine how the geometry, the height and the concentration of corrugations
can alter the depletion potential. 
Although numerical simulations are a powerful tool to investigate the effects of roughness on the 
depletion mechanism, they cannot be efficiently performed for the evaluation of the effective interaction 
in a wide range of parameters characterising the particle corrugation. 
For instance, Ref.~\cite{Schweizer} deals only with hemispherical roughness which appears
to be strongly correlated, whereas in Ref.~\cite{mariolina_2016} a single geometry and a single 
value of the dimension of the particles which cover the colloidal surface are considered.
Furthermore these approaches are not suitable to give a quick estimate of the properties of the potential
and of aggregation, which is critical for an experimentalist interested in the behaviour of colloidal suspension
in presence of entropic interactions.

In this paper, we develop an approximate theoretical approach for the evaluation of the depletion interaction
between two rough spheres. The proposed model is deliberately simple, so to provide analytical expressions for 
the effective potential in the two limits of fully uncorrelated roughness or in the presence of strong 
repulsion between defects on the surface. Within the limits of the theory, we show that 
this approximation is able to capture the essential features of the effective interaction in a significant
range of physical parameters. The model is described and analysed in Section \ref{sec:model}, where the limits of applicability  
to physical systems are also discussed. The results are compared with the recent simulations~\cite{mariolina_2016} 
in Section \ref{sec:results}. In the same Section, the dependence of the effective interaction on the size of the surface 
roughness and on its geometry are investigated. The implications of the different potential shapes
on the efficiency of particle aggregation are also discussed. 

\section{The model}
\label{sec:model}
Let us consider two hard colloidal spheres of radius $R$, whose surface is divided in
patches of area $A \ll R^2$.  Each patch can accommodate at most one bump of height $\epsilon$,
which represents the roughness.
The projection of each bump on the surface of the colloid is a circle of radius $a$,
so the bumps can be viewed as spheres, cylinders, hemispheres and so on.
To reduce the number of parameters of the model, we do not allow
for a statistical distribution in the dimensions and geometry of the bumps.
A relevant parameter of the model is the total number $N$ of bumps on the sphere, 
which defines the dimensionless coverage $c$ as the fraction of the spherical
surface covered by bumps:
\begin{equation}
c=N\frac{a^2}{4 R^2}. 
\label{eq:covering}
\end{equation}
The two colloidal particles are set at a center-to-center distance $r$ and
immersed in an ideal gas of particles (typically polymers), modelled as spheres of
diameter $\sigma \ll R$, suffering a hard core repulsion with the rough colloidal surface.

Following the classical approach of Asakura-Oosawa~\cite{asa_osava_54,asa_osava_58}, the solvent mediated interaction
between the two corrugated spheres can be obtained evaluating the overlap volumes for each realisation of
the disorder which characterises the roughness of the two spheres.
If the patch is small enough so that the colloidal surface can be considered flat
on the scale of the patch size, it is possible to reduce the difficulty of the evaluation of the overlap volume
resorting to Derjaguin approximation for each patch. For a given realisation of the roughness and for
a center-to-center distance $r$ at which the colloids do not overlap we write:
\begin{equation}
V^{ov}(r)=\sum_{j}V^{ov}_{j}(r),
\label{eq:sum_ov_patch}
\end{equation}
where the sum over $j$ runs over all the patches and  $V^{ov}_{j}(r)$ is the overlap volume
generated by the intersection of the excluded volume of the patch $j$ on one sphere 
with the excluded volume belonging to the corresponding facing patch $j$ on the other sphere.
\newline
Due to thermal motion, the rough colloidal particles immersed in the depletant
approach from different directions and accordingly the effective interaction
results to be the average over all the orientations of the two colloidal particles.
Within our formalism, this average process can be mimicked by performing an average over different
realisations of the disorder. This amounts to compute:
\begin{equation}
\beta v_{eff}(r) =- \log \Big\langle e^{-\beta v(r)} \Big\rangle,
\label{eq:average0}
\end{equation}
where angular brackets denote a statistical average over uncorrelated disorder on the
two spherical surfaces.

By taking the colloid diameter $2R$ as a length unit, our model is defined by five parameters:
the bump dimensions $\epsilon$ and $a$ and the coverage $c$ characterising the sphere roughness;
the size ratio $q$ and the reservoir density $z_p$ characterising the polymer solution.

\subsection{Uncorrelated roughness}
\label{sec:uncorrelated_roughness}
\begin{figure}
\centering
  \includegraphics[width=8cm]{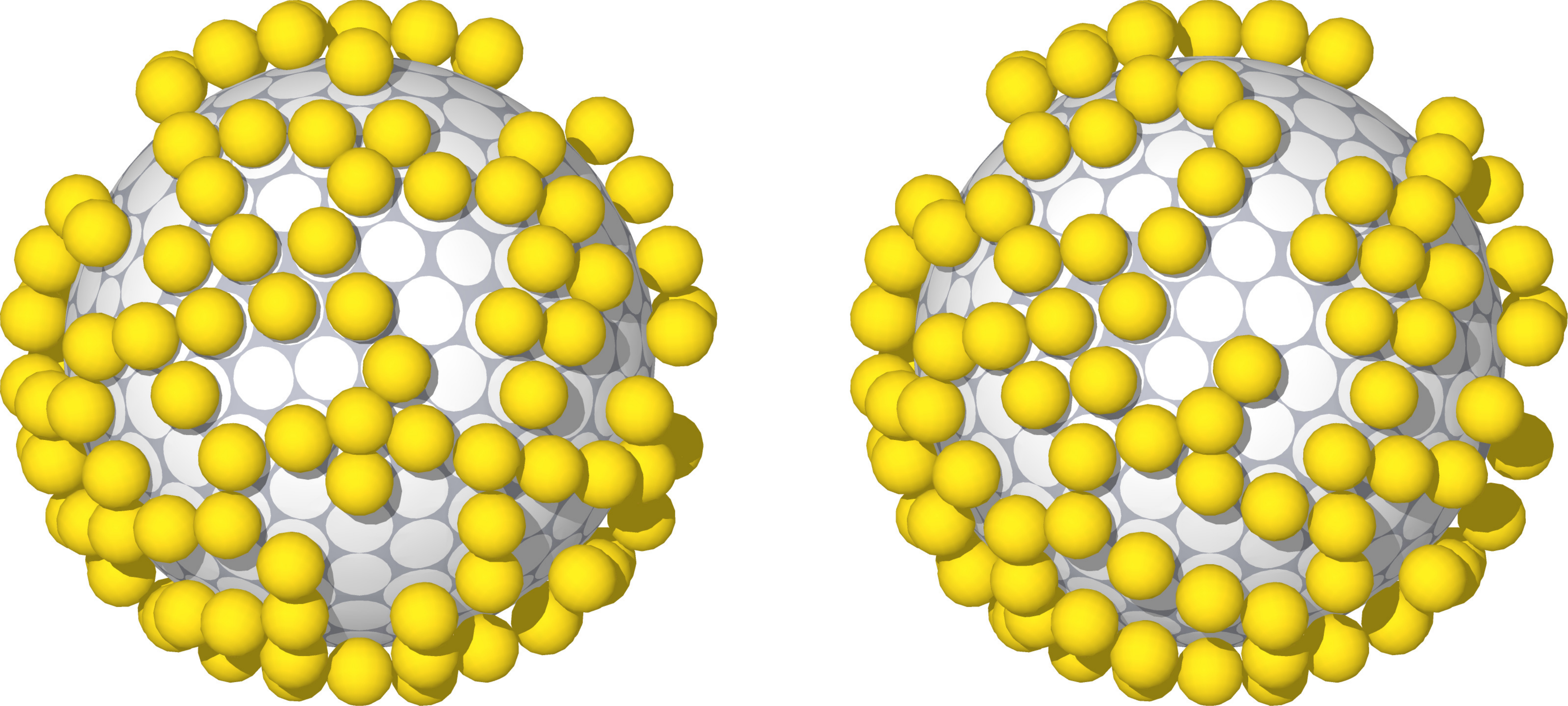}
  \caption{Two facing spheres characterised by uncorrelated roughness. At fixed covering $c$, the patches 
           of radius $a$ are occupied with probability $c$ by spheres of the same radius. 
  }
  \label{fgr:figura_uncorrelated}
\end{figure}

We begin by considering the case of uncorrelated roughness, i.e. when the bumps are located 
randomly on the colloidal surface. 
We divide as showed in Figure \ref{fgr:figura_uncorrelated} the colloidal particle surface 
in circular patches of radius $a$ and area $A=\pi a^2$. 
Each patch has a probability $c$ to host a bump and, according to Derjaguin approximation, 
the overlap volume $V^{ov}(r)$ can be written as in Eq.~(\ref{eq:sum_ov_patch}). 
Then the average of the Boltzmann weight over different realisations of disorder is
given by:
\begin{align}
\Big\langle e^{-\beta v(r)} \Big\rangle &= \prod_j \biggl[ c^2 \chi_j^{11} e^{z_p V_j^{11}}+ 
2c(1-c)\chi_j^{10}e^{z_p V_j^{10}} \notag \\
&\qquad\qquad\qquad\qquad\qquad \qquad + (1-c)^2\chi_j^{00}e^{z_p V_j^{00}}\biggr],
\label{eq:average}
\end{align}
where the index $j$ labels the patches on each sphere; the pair of indices $(11)$, $(10)$ and $(00)$ label the three
possibilities of having a bump on the facing patches of both spheres, having a bump only on one patch and having no
bumps in both patches; $\chi_j^{\mu\nu}=0,1$ according whether the configuration $(\mu\nu)$ is possible (i.e. does not
violate the hard core constraint) for the patches labelled by $j$; $V_j^{\mu\nu}$ is just a geometrical quantity,
which depends on the shape of the bumps and on the distance between the two facing patches $j$,
defining the overlap volume of the depletion layers for the patch configuration $(\mu\nu)$.
Taking the logarithm of Eq. (\ref{eq:average}) and evaluating the sum in terms of an integral over the surface of
the colloidal particle, we get
\begin{align}
\beta v_{eff}(r) &=-\frac{2\pi R^2}{A}\,\int_0^\pi {\rm d}\theta\,\sin\theta \,
\log \biggl[ c^2 \chi^{11} e^{z_p V^{11}} \notag \\ 
& \qquad \qquad + 2\,c\,(1-c)\chi^{10}e^{z_p V^{10}}
+(1-c)^2\chi^{00}e^{z_p V^{00}} \biggr], 
\label{eq:average_theta}
\end{align}
where, at fixed center-to-center distance $r$, the factors $\chi^{\mu\nu}$ and $V^{\mu\nu}$ 
both depend on the angular coordinate $\theta$.
\newline
The relation which links, at fixed $r$, the patch-to-patch distance $h_r$ with $\theta$
\begin{equation}
h_r(\theta)=r -2\,R\,\cos\theta
\label{eq:distpatch}
\end{equation}
allows to convert the angular integral in (\ref{eq:average_theta}) into an integral over $h$, leading to our final expression 
for the average effective interaction:
\begin{align}
\beta v_{eff}(r) &= -\frac{R}{a^2}\,\int_{r-2R}^{2\epsilon+\sigma}  {\rm d} h \,
\log \biggl[ c^2 \Theta (h-2\epsilon) e^{z_p V^{11}} \notag \\
& \qquad \quad + 2\,c\,(1-c)\Theta (h-\epsilon) e^{z_p V^{10}} +(1-c)^2 e^{z_p V^{00}}\biggr], 
\label{eq:final_uncorrelated}
\end{align}
where we used the definition of the factors $\chi^{\mu\nu}$
\begin{equation}
\chi^{\mu\nu}(h)=
\begin{cases}
1 & \text{if $h>(\mu+\nu)\,\epsilon$}\\
0 & \text{elsewhere}
\end{cases},
\notag
\end{equation}
and $\Theta(\cdot)$ is the Heaviside step function.
The overlap volumes $V^{\mu\nu}(h)$ depend on the shape of the bumps representing the surface roughness,
and always vanish for $h>(\mu+\nu)\,\epsilon+\sigma$.
If we assume spherical bumps of diameter $\epsilon=2a$, as in the recent numerical study~\cite{mariolina_2016},
the explicit expressions are:
\begin{align}
V^{11}(h) &= 
\frac{\pi}{6}\left(\epsilon+2\sigma+h\right)\left(2\epsilon+\sigma-h\right)^2  &\epsilon+\sigma<h&<2\,\epsilon+\sigma& \notag \\
V^{10}(h)  &= 
\frac{\pi}{6}\left(\epsilon+\sigma+2h\right)\left(\epsilon+\sigma-h\right)^2  &\sigma<h&<\epsilon+\sigma& \label{eq:v10} \\
V^{00}(h)  &= \pi\frac{\epsilon^2}{4} (\sigma-h) &0<h&<\sigma \notag
\end{align}
For each geometry of the bumps the effective potential is easily obtained 
evaluating numerically the integral in Eq. (\ref{eq:final_uncorrelated}). 

\subsubsection*{Effective potential at $\eta=0$}
When the depletant is absent, i.e. $z_p=0$, the average over the disorder of Eq. (\ref{eq:average0}) can be carried out
analytically. The effective potential is purely repulsive and can be written as:
\begin{equation}
\beta v^{R}_{eff}(r)\!=\!
\begin{cases}
+\,\infty & r<2R \\
\!-\!\left(\!\frac{2R}{a}\!\right)^2\!\Bigl[\!\left(1\!+\!\frac{\epsilon}{2R}\!-\!\frac{r}{2R}\right)\!f(c)\!+\!\frac{\epsilon}{2R}g(c)\!\Bigr] & 2R\!\le\! r\! <\! 2R\!+\!\epsilon\\
\!-\!\left(\!\frac{2R}{a}\!\right)^2\! \left(1\!+\!\frac{\epsilon}{R}\!-\!\frac{r}{2R}\!\right)\!g(c) & \!2R\!+\!\epsilon\! \le\! r\! <\!2R\!+\!2\epsilon \\
0 & r \ge 2R+2\epsilon 
\end{cases}
\label{eq:corr_rep}
\end{equation}
where $f(c)=\log(1-c)$ and $g(c)=\log\sqrt{1-c^2}$.
The potential shows a simple behaviour: it vanishes for $r\ge2R+2\epsilon$ while for $2R<r< 2R+2\epsilon$ it is
formed by two straight lines with different slopes, proportional to $f(c)$ and $g(c)$, joined at $r=2R+\epsilon$.     

\subsection{Correlated roughness}
\label{sec:correlated_roughness}
\begin{figure}
\centering
  \includegraphics[width=8cm]{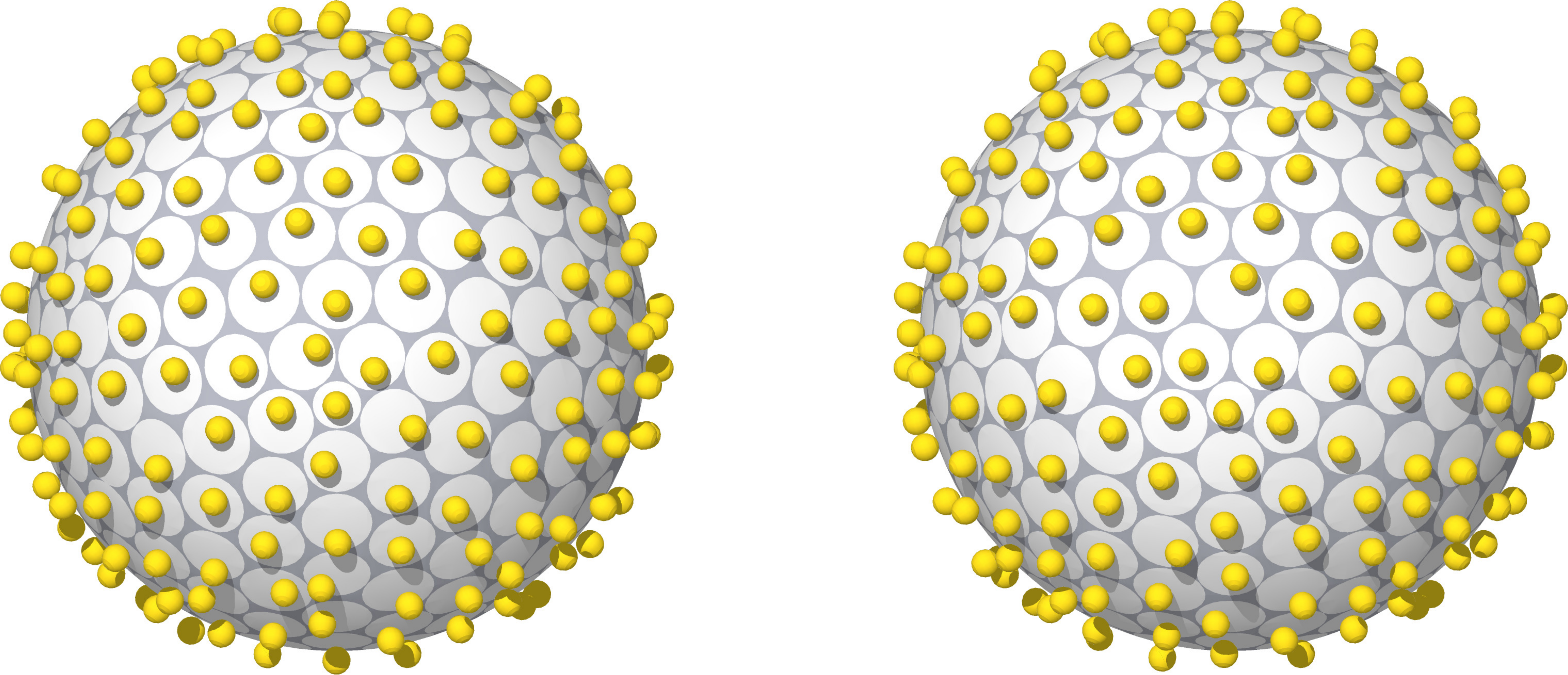}
  \caption{Two facing spheres characterised by correlated roughness. At fixed covering $c$, all the patches
           of radius $a/\sqrt{c}>a$ are occupied a sphere of radius $a$ located randomly
           within the patch.
  }
  \label{fgr:figura_correlated}
\end{figure}

In the experimental realisations of surface roughness, bumps are often electrically charged ~\cite{mariolina_2016}.
The occurrence of repulsive interactions can favour a more homogeneous covering
of the particle surface: In this situation the bumps are randomly distributed on the colloid,
but at the same time long range fluctuation are inhibited (see Figure \ref{fgr:figura_correlated}). This behaviour is expected to be more relevant 
at high covering, when the distance of bumps reduces.
\newline
In order to model this situation, we require that each patch of area $A$ accommodates exactly one
bump, whose projection on the patch surface is $\pi a^2<A$.
The bump is placed randomly within the patch and $A$ is determined by enforcing 
the surface covering condition $N A =4\pi R^2$ ($N$ is both the number of patches and 
of bumps)  which leads through Eq. (\ref{eq:covering}) to
\begin{equation}
A=\frac{\pi a^2}{c}.
\end{equation}
Once again for the calculation of the overlap volumes we resort to Derjaguin approximation 
for each patch and the statistical average in Eq. (\ref{eq:average0}) is now performed 
over the random location of the bump in the patch.
In the case of spherical bumps Eq. (\ref{eq:average}) is replaced by:
\begin{equation}
\Big\langle e^{-\beta v(r)} \Big\rangle = \prod_j  \biggl[ \frac{1}{A^2} \int \mathrm{d} {\bm r}_1\,\mathrm{d} {\bm r}_2 \, 
\Theta \left(d_{h_j}(|{\bm r}_1-{\bm r}_2|)- \epsilon \right) \,e^{z_pV_j(|{\bm r}_1-{\bm r}_2|)}\biggr],
\notag
\end{equation}
where $h_j$ is the distance between two facing patches (\ref{eq:distpatch}); the couple of 
two-dimensional integrals are extended to the surface of a single patch;
\begin{equation}
d_{h_j}(|{\bm r}_1-{\bm r}_2|)=\sqrt{|{\bm r}_1-{\bm r}_2|^2+(h_j-\epsilon)^2} 
\label{eq:dhj}
\end{equation}
is the surface-to-surface
distance between the two bumps hosted by the facing patches $j$ and located on the 
two colloidal particles at position ${\bm r}_1$ and ${\bm r}_2$; 
$V_j$ is the overlap volume between the depletion layers. 
Following the same steps which lead to Eq. (\ref{eq:final_uncorrelated}) we get:
\begin{align}
\beta v_{eff}(r) &= -\pi\frac{R}{A}\,\int_{r-2R}^{2\epsilon+\sigma} \mathrm{d}h 
\, \log \biggl[ \frac{1}{A^2}  \int \mathrm{d} {\bm r}_1\,\mathrm{d} {\bm r}_2 \notag \\ 
& \qquad \qquad \qquad \Theta\bigl(d_h(|{\bm r}_1-{\bm r}_2|)-\epsilon\bigr)\,e^{z_pV(|{\bm r}_1-{\bm r}_2|,h)} \biggr],
\label{eq:average1-1}
\end{align}
where $d_h$ has been defined above in Eq. (\ref{eq:dhj}).
The overlap volume can be expressed as the sum of two terms:
\begin{equation}
V(|{\bm r}_1-{\bm r}_2|,h)=2 \, V^{10}(h) + V^{11}\bigl(d_h(|{\bm r}_1-{\bm r}_2|)\bigr),
\label{eq:volumi}
\end{equation}
where $V^{10}$ does not depend on the difference ${\bm r}_1-{\bm r}_2$ and coincides with Eq. (\ref{eq:v10}), 
whereas 
\begin{equation}
V^{11}(d) = 
\frac{\pi}{12}\,\left( 2\epsilon +2\sigma+d\right)\left(\epsilon +\sigma-d\right)^2 \qquad \epsilon<d<\epsilon+\sigma
\notag
\end{equation}
Substituting the expression for the overlap volume of Eq. (\ref{eq:volumi}) in Eq. (\ref{eq:average1-1}), 
we obtain our final result for the effective potential in the case of short-range correlated roughness:
\begin{equation}
\beta v_{eff}(r) = -\frac{R}{s^2} \biggl[2 \,z_p \int_{r-2R}^{\epsilon+\sigma} \mathrm{d}h \, V^{10}(h) + 
\int_{r-2R-\epsilon}^{\epsilon+\sigma} \mathrm{d} \xi\, \log \mathcal{K}(\xi) \biggr],
\label{eq:pot_corr_fin}
\end{equation}
where in the second integral on the r.h.s. the change of variable $\xi=h-\epsilon$ has been performed.
The function $\mathcal{K}(\xi)$ is defined:
\begin{equation}
\mathcal{K}(\xi)=\frac{1}{\left(\pi s^2\right)^2}  \int_{|{\bm r}_1|<s} \mathrm{d} {\bm r}_1 \int_{|{\bm r}_2|<s} \mathrm{d} {\bm r}_2  \,
f\left(|{\bm r}_1-{\bm r}_2|;\xi\right),
\label{eq:k_xi}
\end{equation}
where 
\begin{equation}
f(r;\xi)=\Theta \left(\sqrt{r^2+\xi^2}-\epsilon\right)\,e^{z_p \,V^{11}\bigl(\sqrt{r^2+\xi^2}\bigr)} 
\label{eq:defin_fk}
\end{equation}
and $s=a/\sqrt{c}$ is the radius of the patch.
The numerical evaluation of the integrals (see the Appendix for details) provides the effective potential. 

\subsection{Approximations and range of parameters}

The expressions for the effective potential obtained above within the AO framework (i.e. ideal depletant) 
mainly rely on the patch-to-patch evaluation of the overlap volume 
coupled with Derjaguin approximation. 
In addition, we assume that the size and the geometry of the bumps 
are the same for each bump and we do not allow for multiple occupancy of the patches.

Regarding the geometry of the roughness, our approach does not allow 
to model the bumps as objects characterised by a substantial curvature: 
due to the patch-to-patch approximation only a small fraction of the possible overlap volume
would be taken into account. 

The application of Derjaguin approximation requires some care.
First of all the bump size must be much smaller than the colloidal particle size
($a \ll R$), because the effective potential 
is evaluated by decomposing the spherical surface into small patches whose curvature is neglected. 
We also remark that the condition $a \ll R$ guarantees, in the case of uncorrelated roughness, 
that the spherical surface is covered by a large number of patches. 
We expect that our model for correlated roughness works well only at intermediate and high coverings, 
because at small coverings the patch surface becomes large with respect to the spherical surface 
and Derjaguin approximation becomes inaccurate. 
\newline
Furthermore, Derjaguin approximation 
neglects the occurrence of the interstices between two facing bumps
induced by the curvature of the colloidal particle. This circumstance
may lead to a substantial overestimation of the depletion effects 
whenever the depletant diameter $\sigma$ violates the inequality: 
\begin{equation}
\sigma>\frac{4A}{\pi R},
\notag
\end{equation}
where $A$ is the surface of the patch.
\newline
We prevent the overlap between the depletion layers around non nearest neighbour bumps
further requiring that $\sigma< 2a$.

Finally we introduce a condition which relates the parameters $a$ and $\epsilon$.
The present approach neglects the overlap between the excluded volumes belonging to
non-facing patches of the two colloids. 
It is possible to give a rough estimate showing that the effect of the neglected
volume is not relevant if the height of the bump $\epsilon$ is of the same order of magnitude or smaller than the 
radius $a$ of the projection of the bump on the surface of the colloid.

Summarising, we expect that, in the case of spherical bumps ($\epsilon=2a$),
our model provides reliable results when the parameters fulfil the following conditions:
\begin{equation}
\epsilon \ll 1 \qquad \quad \sigma < \epsilon \qquad \quad \sigma > 2\epsilon ^2,
\notag
\end{equation}
where the lengths are measured in units of the colloid diameter.

\section{Results}
\label{sec:results}
The numerical evaluation of the integrals in Equations (\ref{eq:final_uncorrelated}) and (\ref{eq:pot_corr_fin}) 
provides the solvent-mediated potential between two spheres whose surface roughness is characterised 
by the geometry of the bumps and the five parameters introduced above. 
Since the space of the parameters is relatively large, we decided to investigate the behaviour 
of the potential by varying one parameter at a time.
Furthermore, within the AO framework adopted in this work, the polymer packing fraction 
appears only as a multiplicative factor of the overlap volume (see Eq. (\ref{eq:ao_generale})). 
Therefore the dependence of the effective potential on $\eta$ is monotonic: 
the larger is the density of depletant, the more attractive is the potential.

\begin{figure}
\centering
  \includegraphics[width=8cm]{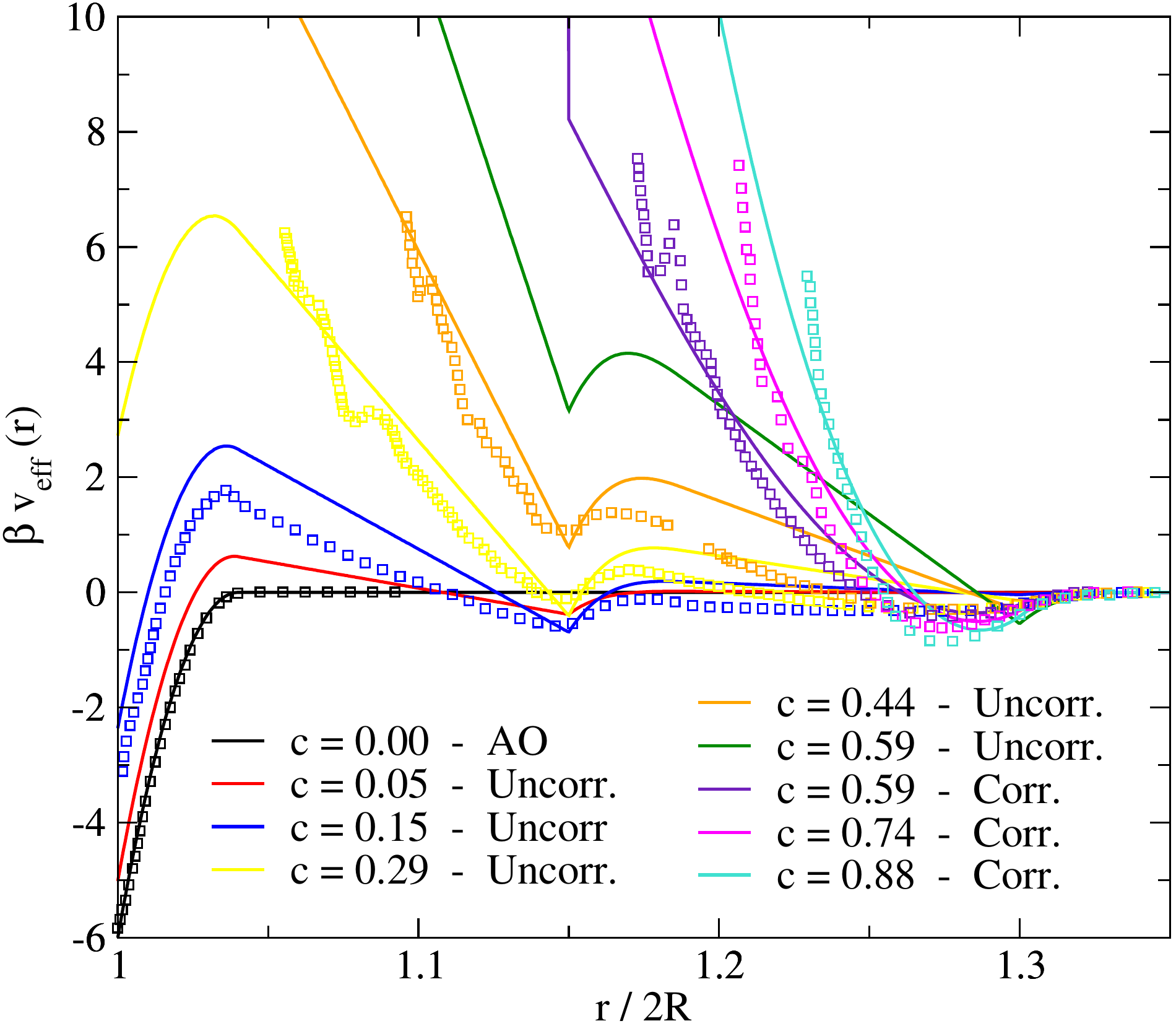}
  \caption{Lines: Effective potential between two rough hard spheres (spherical bumps $\epsilon=2a$)
           as a function of center-to-center separation.
           The packing fraction and the diameter of the depletant are $\eta =0.16$ and $\sigma/2R=0.04$
           respectively. The diameter of a spherical bump is $\epsilon/2R=0.15$, whereas the covering 
           is varied as shown. The curves are evaluated by means of Eq. (\ref{eq:final_uncorrelated})
           in the case of uncorrelated covering ($c \le 0.59$) and by means of Eq. (\ref{eq:pot_corr_fin})
           in the case of correlated covering ($c \ge 0.59$). 
           Points: Data from the MC simulation of Ref.\cite{mariolina_2016}}
  \label{fgr:mariolina_nostriloro}
\end{figure}

Figure \ref{fgr:mariolina_nostriloro} shows the effective potential between two hard spheres 
whose surface is decorated by spherical bumps for different values of the covering $c$ and
the same parameters adopted in the recent simulation\cite{mariolina_2016}. 
We first used the uncorrelated model of roughness 
as defined by Eq. (\ref{eq:final_uncorrelated}) for covering $c\le0.59$. In this case, 
the minimum distance between the two colloidal particles is $2R$ and the potential vanishes for $r > 2R+2\epsilon+\sigma$.
Figure \ref{fgr:mariolina_nostriloro} highlights that the presence of surface roughness decreases the depth of the AO attractive minimum 
of the potential already at small surface coverings. 
By increasing the covering, the potential becomes more and more repulsive except at distances $r\simeq 2R+2\epsilon$,
where an attractive minimum develops due to the presence of a depletion layer around the bumps.
The potential resembles the superposition of an attractive depletion contribution, arising from the overlap 
between $00$, $10$ and $11$ configurations present at center-to-center distances $r\simeq 2R$, $r\simeq 2R+\epsilon$ and 
$r\simeq 2R+2\epsilon$, and the repulsive contribution in Eq. (\ref{eq:corr_rep}) present at all distances.
\newline
By comparing our analytical expression (\ref{eq:final_uncorrelated}) with recent 
numerical simulations\cite{mariolina_2016}, also shown in Fig. \ref{fgr:mariolina_nostriloro},
it appears that the model of random roughness we developed in Section \ref{sec:uncorrelated_roughness}
captures the overall shape of the effective interaction and the agreement is quantitative up to a covering $c=0.44$. 
For higher values of $c$ a qualitative change in the simulation results occurs, suggesting 
that some other effect becomes relevant in the regime of high coverings:
The interaction is much more repulsive at short
distances and becomes attractive at $r\simeq 2R + 2\epsilon$.
The numerical results at intermediate/high coverings (i.e. for $c>0.5$) can be well reproduced by the
the approach developed in Section \ref{sec:correlated_roughness}, where bumps are assumed to be distributed 
in a more uniform way on the surface of the colloidal particle. 
In this case, the range of the potential is the same as for the uncorrelated model, but the distance of 
closest approach is increased to $2R+\epsilon$. The potential is monotonic in the repulsive region 
until a minimum is reached at distances $r\simeq 2R+2\epsilon$ due to the presence of the depletion layer
on two facing bumps belonging to the two colloidal particles. 
The good agreement with the numerical data in Figure \ref{fgr:mariolina_nostriloro} 
shows that the model of correlated bumps provides a faithful representation of the simulation results,
suggesting that the procedure adopted in Ref.\cite{mariolina_2016} for modelling the surface roughness induces
repulsive correlations among bumps already at intermediate coverage.
\newline
Having determined the form of the effective interaction, we can 
discuss the implications of surface roughness on the tendency towards aggregation of the 
two colloidal particles by evaluating 
the reduced second virial coefficient \bibnote{The reduced second virial coefficient 
is defined as the ratio of the second virial coefficient $B_{2}$ and the second virial coefficient $B_{2}^{\mathrm{HS}}$ 
of a hard sphere system with particles of radius $R$:
\begin{equation}
B_{2}^{*}=\frac{B_{2}}{B_{2}^{\mathrm{HS}}}=
1-\frac{3}{8R^3}\int_{2R}^{\infty}\mathrm{d}r\,r^2\left(e^{-\beta v_{eff}(r)}-1\right).
\end{equation}
According to Noro-Frenkel extended law of corresponding states\cite{noro_extended_2000}, 
for hard sphere fluids characterised by short ranged interaction potentials, the second virial coefficient 
assumes a value of about $-1.6$ at the critical point, independently on the particular
form of the interaction. Furthermore its value remains constant in a relatively 
large density range across the critical point.} $B_2^{*}$.
In the case of bumps placed at random on the surface, $B_2^{*}$ changes from $-3.5$ for smooth spheres
to $-0.7$ for $c=0.05$, where aggregation is inhibited by the roughness. Clearly, by increasing
the coverage, the colloidal particles behave almost as hard spheres, at least for the choice of
parameters investigated here.

\begin{figure}
\centering
  \includegraphics[width=8cm]{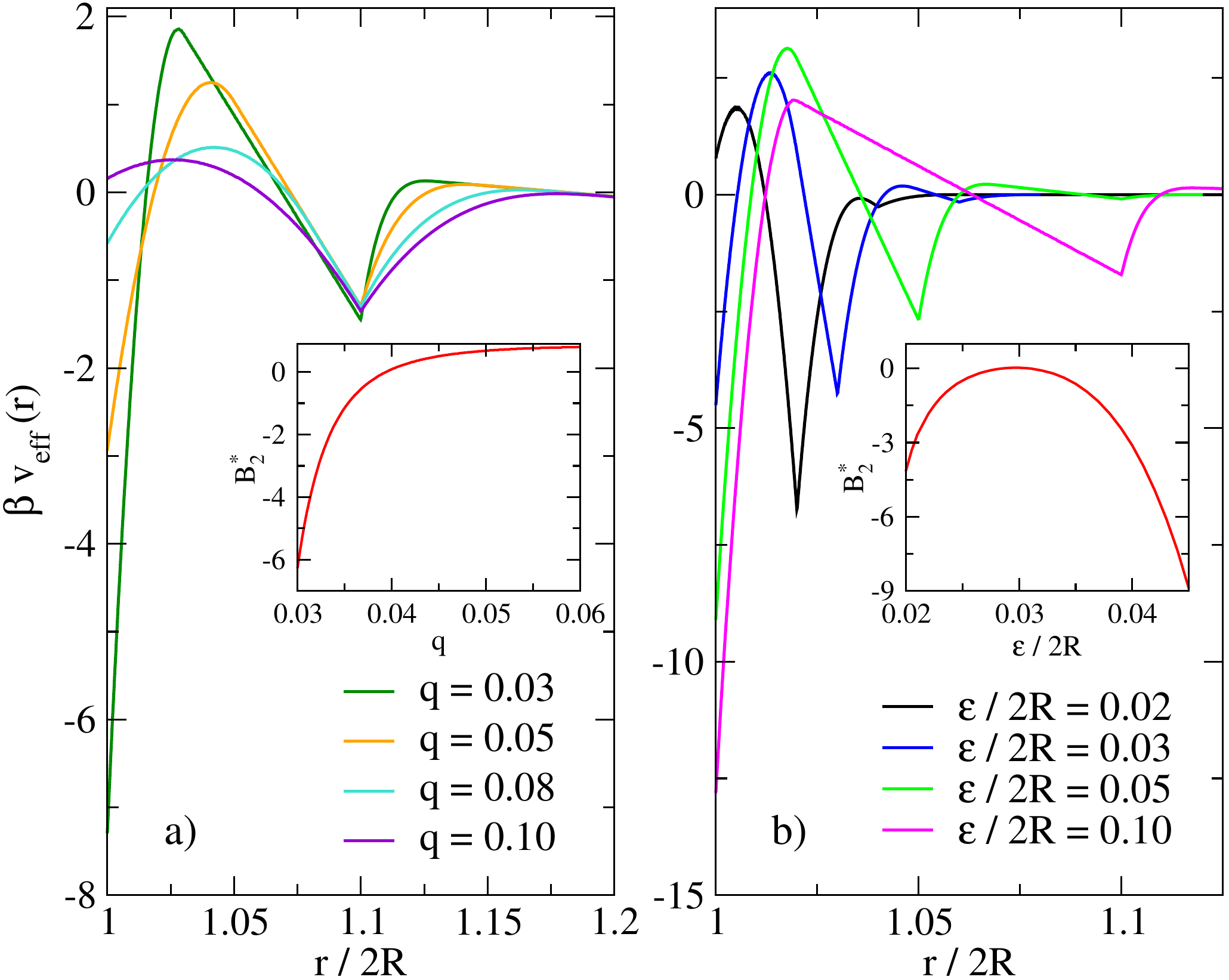}
  \caption{Effective pair potential potential as a function of the center-to-center separation between two rough spheres 
           (spherical bumps) evaluated by means of Eq. (\ref{eq:final_uncorrelated}). In both panels 
           $\eta=0.21$ and $c=0.11$. Panel a): $\epsilon/2R =0.1$, $q$ ranges from $0.03$ to $0.1$ as shown.
           The inset shows the reduced second virial coefficient $B_2^{*}$ as a function of $q$.
           Panel b): $q=0.02$, $\epsilon/2R$ ranges from $0.02$ to $0.1$ as shown. The inset shows the  
           reduced second virial coefficient $B_2^{*}$ as a function of $\epsilon$.}
  \label{fgr:varsigmavareps}
\end{figure}

Figure \ref{fgr:varsigmavareps} compares the behaviour of the effective potential between two hard spheres
decorated by spherical bumps at different values of the ratio $\sigma/\epsilon$ between the depletant diameter and
the height of the bumps.
In panel a) the value of $\epsilon$ is hold fixed whereas $\sigma$ varies. When $\sigma$ is small ($q=\sigma/2R=0.03$)
a deep short-ranged minimum of the potential develops at contact and the reduced second virial coefficient is 
 negative ($B_2^{*}\simeq -8$).
The increase of $\sigma$ reflects in an increase of $B_2^{*}$, becoming 
positive at $q\simeq0.04$ (see the inset). This result depends on the fact that the depletion potential becomes 
weaker at larger size ratio $q$, as in the case of smooth spheres (\ref{eq:contAO}),
while the repulsive contribution to the interaction does not change.
In panel b) the comparison is carried out at fixed $\sigma$. When $\epsilon=\sigma$ the potential 
is repulsive at contact, but a second attractive minimum of the order of $7 k_{\mathrm{B}}T$ appears at $r=2R+\epsilon$
and $B_2^{*}\simeq -4$. This minimum becomes less pronounced for larger $\epsilon$ while the AO attraction at contact develops. 
For this choice of the parameters $B_2^{*}$ shows a peculiar non monotonic behaviour illustrated in the inset: At $\epsilon/2R\simeq0.03$ 
the reduced second virial coefficient is close to zero, whereas at $\epsilon/2R\simeq0.05$ it becomes strongly negative again. 
This unexpected behaviour can be attributed to the fact that, in the case of spherical bumps,
by increasing $\epsilon$ also the bump radius $a={\epsilon}/{2}$ increases and, at fixed coverage, the number of 
corrugations reduces, exposing larger available portions of the underlying particle surface to the depletion mechanism. 

\begin{figure}
\centering
  \includegraphics[width=8cm]{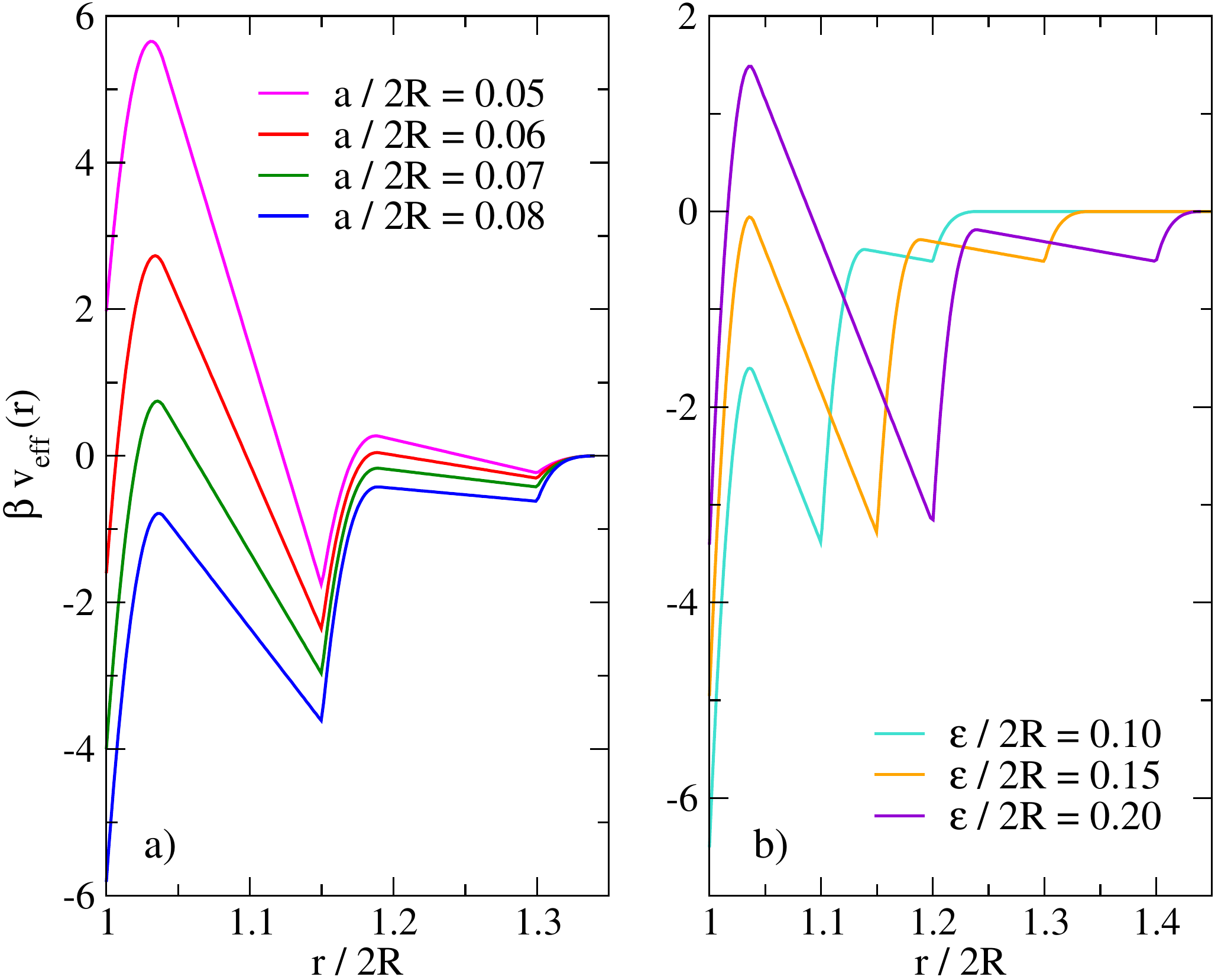}
  \caption{Effective potential between two rough spheres as a function of the center-to-center separation evaluated 
           using Eq. (\ref{eq:final_uncorrelated}). The spherical surface is decorated with cylinders with basis of 
           radius $a$ and height $\epsilon$.  In both panels $\eta=0.16$, $q=0.04$ and $c=0.11$.
           Panel a): $\epsilon/2R =0.15$, $a/2R$ ranges from $0.05$ to $0.08$ as shown.
           Panel b): $a/2R=0.075$, $\epsilon/2R$ ranges from $0.1$ to $0.2$ as shown.}
  \label{fgr:vara_vareps}
\end{figure}

The geometry of the bumps significantly affects the shape of the depletion potential. Our model
allows to investigate this effect: 
Spherical bumps constrain the height of the surface roughness (defined by the parameter $\epsilon$) and 
the section of each bump (related to the parameter $a$) by the relation $\epsilon=2a$. This limitation is lifted
in the case of cylindrical bumps with radius $a$ and height $\epsilon$, thereby representing a simple model of roughness allowing to 
study the effects of these two parameters separately. In Figure \ref{fgr:vara_vareps} we display some representative result
for such a choice. 
Panel a) shows that, at constant covering $c$ and height $\epsilon$, the roughness is more effective when the surface is
covered by a large amount of small bumps than a small number of corrugations with a large surface. 
Instead, when the number of bumps is constant, the potential is more repulsive in the case of larger $\epsilon$, as can be 
seen in panel b).

\begin{figure}
\centering
  \includegraphics[width=8cm]{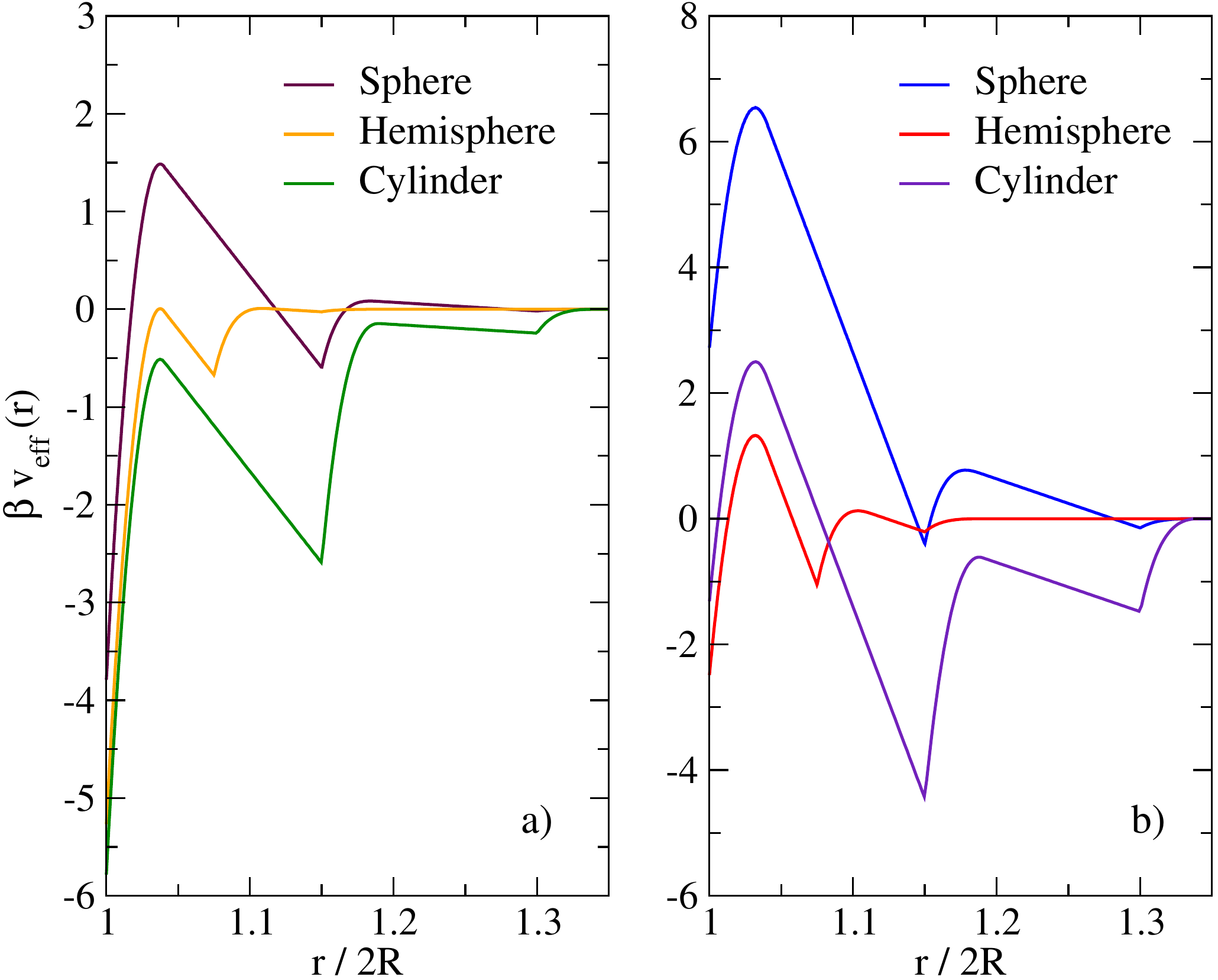}
  \caption{Effective potential between two rough spheres as a function of the center-to-center separation evaluated 
           using Eq. (\ref{eq:final_uncorrelated}). The spherical surface is decorated with spheres ($\epsilon/2R=a/R=0.15$), 
           cylinders ($\epsilon/2R=0.15$, $a/2R=0.075$) and hemispheres ($\epsilon/2R=a/2R=0.075$). 
           In both panels $\eta=0.16$, $q=0.04$.
           Panel a): $c =0.10$. Panel b): $c=0.29$.}
  \label{fgr:confronto_geo}
\end{figure}

Figure \ref{fgr:confronto_geo} compares the potential obtained with bumps characterised by different geometries
in the case of uncorrelated roughness for two values of covering. The spheres and the cylinders have 
the same height, while the hemispheres are obtained dividing the spheres into two equal parts.
The spherical and hemispherical geometry proves to be more effective in suppressing the depletion interaction with 
respect to the cylindrical geometry at the same covering. 
This happens because the curvature of spherical bumps reduces the overlap volume with respect to the flat surface of cylinders.
It is interesting to note that, in the case of cylindrical bumps, the potential develops a quite deep attractive minimum
at $r=2R+\epsilon$ caused by the large overlap volume arising when a bump on one sphere faces a portion of smooth surface 
on the other particle. 

\section{Conclusions}

We introduced an analytical model able to capture the effects of surface roughness on the depletion 
mechanism. Comparison with available simulations shows that this, admittedly very schematic, parameter free model 
allows to quantitatively reproduce the main features of the effective interaction between two spherical colloids 
in the presence of random surface corrugation. 
In order to reduce the problem to an analytically tractable model, we introduced several approximations. The two most 
relevant assumptions are: 
\begin{itemize}
\item 
The uncorrelated character of the corrugations. The surface of the particle 
has been divided in patches of the same size of the corrugation and a sort of ``Ising variable'' has been defined on each 
patch, representing the occupancy of each patch. This procedure neglects the correlations between the presence of bumps on 
different patches, and is therefore appropriate in the limit of small coverage. 
We also introduced an alternative model, where surface corrugation is strongly correlated at short distances, as expected 
when a repulsive interaction between bumps is present. This model has been shown to accurately describe the case of high 
coverage in the presence of charged adsorbed particles. 
\item 
Another important approximation we introduced refers to the calculation of the overlap volume between 
facing patches on the two particles. In the spirit of mean field theories, we disregarded the 
effects induced by the presence of two nearby bumps on the overlap volume, thereby neglecting correlations between 
nearest neighbour corrugations.
\end{itemize}

We found that, as expected, surface roughness deeply
inhibits the depletion effects, strongly reducing the tendency towards aggregation of the colloidal 
particles. These findings confirm that irregularities in the particle surface play a key role in 
the properties of colloidal suspensions. As a general rule, the effects induced by surface roughness appear to be more
relevant when the height of the corrugations is of the same order of the size of the depletant.  
Several parameters are necessary to describe, even approximately, the extent of surface roughness: 
The height, width, number and shape of the bumps on each particle often induce competing effects on
the depletion potential. 
\newline
The availability of an analytical model may be extremely useful to 
estimate the effects of a specific surface roughness on the overall features of the effective interaction,
even at a semi-quantitative level, without resorting to numerical simulations.





\section*{Appendix}

Exploiting the symmetry of the problem it is possible to reduce the expression in Eq. (\ref{eq:k_xi}) 
in order to perform a numerical evaluation.
\newline
The first step is to perform the change of variables ${\bm r}_2={\bm r}_1 - {\bm r}$, obtaining
\begin{equation}
\mathcal{K}(\xi)=\frac{2}{\pi s^4}  \int_{0}^{2s} \mathrm{d}r\,r\,f(r;\xi) \int_{\mathcal{D}} \mathrm{d} {\bm r}_1
\notag
\end{equation}
where the domain of the integral is $\mathcal{D}=\left\{{\bm r}_1:|{\bm r}_1|<s\land |{\bm r}_1-{\bm r}|<s\right\}$ and
we have taken advantage of the central symmetry of the integral over ${\bm r}$.
The integral over ${\bm r}_1$ represents the surface of intersection between two disks of radius $s$ at 
distance ${\bm r}$. The straightforward analytical solution allows to write:
\begin{equation}
\mathcal{K}(\xi)=\frac{4}{s^2}\int_{0}^{2s} \mathrm{d}r\,r\,
\Theta \left(\sqrt{r^2+\xi^2}-\epsilon\right)\,e^{z_p \,V^{11}\bigl(\sqrt{r^2+\xi^2}\bigr)} 
\mathcal{I}\biggl(\frac{r}{2s}\biggr)
\notag
\end{equation}
where we used the definition of $f(r;\xi)$ in Eq. \ref{eq:defin_fk} and
\begin{equation}
\mathcal{I}\left(x\right)=\frac{1}{\pi}\biggl[\arccos\left(x\right)-x\,\sqrt{1-x^2}\biggr]
\notag
\end{equation}
is a function which is non zero for $x\in [0:1]$.
If we replace the variable of integration $r$ with $t=\sqrt{\xi^2+r^2}$ we obtain an 
expression which can be easily integrated numerically
\begin{equation}
\mathcal{K}(\xi)=\frac{4}{s^2}  \int_{ \mathrm{Max}\bigl[\epsilon;\xi\bigr] }^{+\infty} \mathrm{d} t 
\, t \,e^{z_p \,V^{11}(t)} \, \mathcal{I}\left(\frac{1}{2s}\sqrt{t^2-\xi^2}\right).
\notag
\end{equation}




\renewcommand\refname{Notes and references}

\bibliographystyle{rsc} 
\providecommand*{\mcitethebibliography}{\thebibliography}
\csname @ifundefined\endcsname{endmcitethebibliography}
{\let\endmcitethebibliography\endthebibliography}{}

\end{document}